\documentstyle[12pt]{article}
\newcommand{\tr}[1]{\,{\rm tr}\,#1\,}

\begin{document}
\def\newmathop#1{\mathop{#1}\limits}
\def\osim#1{\displaystyle\newmathop{#1}_{\lambda\to 0}}
\def\linfty#1{\displaystyle\newmathop{#1}_{s\to \infty}}
\def\rr#1{\displaystyle\newmathop{#1}_{~~s \to \infty ,~ t~fixed~~}}
\title{
\begin{flushright}
{\small CVV-203-95 \\
SMI-06-95 }
\end{flushright}
\vspace{2cm}
 The Master Field for Large $N$ Matrix Models
\\ and \\ Quantum Groups }
\author{L.Accardi ,$~~$
I.Ya.Aref'eva \thanks{Steklov Mathematical Institute,
Vavilov 42, GSP-1, 117966, Moscow, e-mail: Arefeva@arevol.mian.su},$~~$
S.V. Kozyrev \thanks{Steklov Mathematical Institute,
Vavilov 42, GSP-1, 117966, Moscow, }$~~$
and $~~$ I.V.Volovich \thanks{On leave absence from
Steklov Mathematical Institute,
Vavilov 42, GSP-1, 117966, Moscow}
\\$~$\\
{\it Centro V. Volterra Universit\`a di Roma Tor Vergata}\\
{\it via di Tor Vergata, 00133 -- Roma}\\
}
\date {$~$}
\maketitle
\begin {abstract}
In recent works by Singer, Douglas and Gopakumar and Gross an application of
results of Voiculescu from non-commutative probability theory to constructions
of the master field for large $N$ matrix field theories  have been suggested.
In this note we consider interrelations between the master field and quantum
groups. We define the master field algebra and observe that it is isomorphic
to the algebra of functions on the quantum group $SU_q(2)$ for $q=0$. The
master field becomes a central element of the quantum group Hopf algebra. The
quantum Haar measure on the $SU_q(2)$ for any $q$ gives the Wigner semicircle
distribution for the master field. Coherent states on $SU_q(2)$ become
coherent states in the master field theory.
\end {abstract}
\newpage
\section{Introduction}
\setcounter{equation}{0}

Recently it has been a reveal of interest in the theory of master field
describing the large $N$ limit of matrix models  \cite {Sin}-\cite {AAV}.
Much of the interest in large $N$ expansions is motivated by the desire
to find reliable methods for analyzing the dynamics of QCD  \cite {tH}.
QCD at  large $N$ provides phenomenologically an appealing picture of
strong interactions. In fact $\frac{1}{N}$ provides the only known
expansion parameter which can be used in calculations of hadronic properties
\cite {Ven,Wit}. Different methods has been proposed for finding the  large $N$
limit of various theories  \cite {BIPZ}-\cite {Yaffe}.
However all of them are effective only for low-dimensional models.
Mathematical structures appearing
in the  large $N$ limit are related with free independent algebras
and so called free or Boltzmannian Fock space  \cite {Haan,Voi} deserve a
thorough study. An investigation of these questions have been performed
in non-commutative probability theory  \cite {Voi91,ALV}. A model of quantum
field theory with interaction in the free  Fock space has been considered
in \cite {AAV}.

The master field $\Phi$ for the Gaussian
matrix model is defined by the relation
\begin {equation} 
                                                          \label {mc}
\lim _{N\to\infty} \frac{1}{Z_{N}}\int \frac{1}{N^{1+k/2}}\tr M^{k}
e^{-S(M)} dM =<0|\Phi ^{k}|0>
\end   {equation} 
for $k=1,2...$ , where the action $S(M)={1\over 2}\tr M^{2}$  and
$M$ is an Hermitian $N\times N $ matrix.

The operator $\Phi= a+a^{*}$  acts in the so called
free or Boltzmannian Fock space with creation and annihilation operators
satisfying the relation
\begin {equation} 
                                                          \label {aa}
aa^{*}=1
\end   {equation} 
The relations  (\ref {mc}), (\ref {aa}) have been obtained in physical
\cite {Haan} and mathematical \cite {Voi} works. It can be interpreted
as a central limit theorem in non-commutative probability theory, for a
review see  \cite {ALV}.
The basic notion of non-commutative probability theory is an algebraic
probability space, i.e a pair $(A,h)$  where $A$ is an algebra
and $h$ is a positive   linear functional on $A$. An example of
the algebraic probability space is given by the algebra of random matrix
with (\ref {mc}) being    non-commutative central limit theorem.
As another example one can consider quantum groups.
Theory of quantum groups have received in the last years a lot of
attention  \cite {Dri}-\cite {Jimbo}. In this case
$A$ is the Hopf algebra of functions on the quantum group and $h$
is the quantum Haar measure.

In this note we discuss relations of theory of the master field
in the Boltzmannian Fock space with quantum groups.
If one has $q$-deformed canonical commutation relations (see for
example  \cite {AV})
\begin {equation} 
                                                          \label {ar}
aa^{*}-qa^{*}a=1
\end   {equation} 
then for $q=0$ one gets the relation (\ref {aa}) in the  Boltzmannian
Fock space.

We will show that in fact one has more. In Sect.2 we define an algebra
describing the master field (the master field algebra) and show that
this algebra is isomorphic to the algebra of functions on the quantum group
$SU_{q}(2)$ for $q=0$.
In fact the master field algebra coincides with the algebra of
so called central elements of quantum group  Hopf algebra.
It is interesting, that the transfer matrix in quantum inverse transform
method  \cite {FRT}  is the central element of Yangian
Hopf algebra. In this sense the transfer matrix is an analog of the master
field.  In Sect.3 we show how the canonical master field algebra is related
with quantum groups. In Sect.4 we demonstrate how the quantum group
methods can be used to perform calculations in the large $N$ limit of
matrix models. Namely, the  quantum
Haar measure on the $SU_q(2)$  for any $q$ gives
the Wigner semicircle distribution for the master field. Coherent states on
$SU_q(2)$ become coherent states in the master field theory.

\section{The Master Field Algebra and $Fun(SU_{q}(2))$}
\setcounter{equation}{0}

The free (or Boltzmannian)  Fock space $F$ over the Hilbert space $H$
is just the tensor algebra
$$F=\oplus_{n=0}^{\infty}H^{\otimes n}.$$
Creation and annihilation operators are defined as
$$a^{*}(f) f_{1}\otimes...\otimes f_{n}=f\otimes f_{1}\otimes...\otimes f_{n}$$
$$a(f) f_{1}\otimes...\otimes f_{n}=<f,f_{1}>\otimes f_{2}\otimes...\otimes
f_{n}$$
where $<f,g>$ is the inner product in $H$.
We shall consider the simplest case  $H=C$. One has the vacuum vector
$|0>$,
\begin{equation}\label{vacuum}
a|0>=0
\end{equation}
and the relations
\begin{equation}\label{aac}
aa^{*}=1,
\end{equation}
\begin{equation}\label{aca}
a^{*}a=1-|0><0|.
\end{equation}
We shall reformulate equations (\ref{vacuum}), (\ref{aac}), (\ref{aca})
in the algebraic form. Let us define an operator
\begin{equation}\label{F}
F=e^{i\phi}|0><0|,
\end{equation}
where $\phi$ is an arbitrary real number. Then from
(\ref{vacuum}), (\ref{aac}), (\ref{aca}) one has the following
relations:
\begin {equation} 
                                                          \label {aF}
\begin{array}{ccc}
aF=0, & aF^{*}=0, & FF^{*}=F^{*}F, \\
aa^{*}=1,  &   a^{*}a+FF^{*}=1.
\end{array}
\end   {equation} 

We call the algebra (\ref{aF})  the master field algebra.
 From equations (\ref{aF}) we get
$$(FF^{*})^{2}=FF^{*}$$
and the operator $FF^{*}$ is an orthogonal projector.

Now let us recall the definition of algebra of functions
$A_{q}=Fun(SU_{q}(2))$  on the quantum group  $SU_{q}(2)$.
The algebra $A_{q}$ is the Hopf algebra with generators
$a, a^{*}, c, c^{*}$ satisfying the relations

\begin {equation} 
                                                          \label {comcon}
\begin{array}{ccc}

ac^{*}=qc^{*}a,   & ac=qca,                 &  cc^{*}=c^{*}c, \\
a^{*}a+cc^{*}=1,  &  aa^{*}+q^{2}cc^{*}=1

\end{array}
\end   {equation} 
where $0<|q|<1$. One can get the relations (\ref{comcon}) as the unitarity
condition
\begin{equation}
                                                           \label{ggcrest}
gg^{*}=g^{*}g=1
\end{equation}
for the matrix $g=(g^{i}_{j})$ in the following canonical form
\begin{equation}
                                                           \label{gmatrix}
g=\left(\begin{array}{cc}a&-qc^{*}\\c&a^{*}\end{array}\right).
\end{equation}

$A_{q}$ is a Hopf algebra with the standard  coproduct,
$$\Delta :A_{q}\to A_{q}\otimes A_{q}$$
$$\Delta (g^{i}_{j})=\sum_{k=0,1}g^{i}_{k}\otimes g^{k}_{j}$$
i.e.
$$\Delta(a)=a\otimes a -q c^{*}\otimes c,$$
$$\Delta(c)=c\otimes a + a^{*}\otimes c,$$
with counit $\epsilon:A_{q}\to C, \epsilon(g^{i}_{j})=\delta^{i}_{j}$,
involution * and antipode $S:A_{q}\to A_{q}, $~$Sg=g^{*}$.

Taking $q=0$ in  (\ref{comcon}) one gets the relations (\ref{aF})
if $F=c$. Therefore the master field algebra (\ref{aF})
is isomorphic to the algebra $A_{0}$ of functions
on the quantum group  $SU_{q}(2)$ for $q=0$.

Let $\sigma$ denote the flip automorphism of $A_{q}\otimes A_{q}$:
$$\sigma(x\otimes y)=y\otimes x$$
for any $x, y \in A_{q}$ and
$$A_{q}\otimes_{sym} A_{q}=\{z\in A_{q}\otimes A_{q}:\sigma(z)=z\}.$$
We say that $x$ is the central element of the bialgebra $A$
if $\Delta(x)\in A\otimes_{sym} A.$
An element $x\in A_{q}=Fun(SU_{q}(2))$ is central if and only if $x$
is a linear combination of characters.

The master  field
$$\Phi=a+a^{*}$$ is the central element of the Hopf algebra $A_{q}$
because
$$\Delta(\Phi)=a\otimes a + a^{*}\otimes a^{*} \in A_{q}\otimes_{sym} A_{q}.$$
Notice that for $q=0$ one has more central elements in $Fun(SU_{q}(2))$
than for $|q|>0$. For example $a$ and $a^{*}$ are central elements
since
$$\Delta(a)=a\otimes a.$$

The bosonization of the quantum group $SU_{q}(2)$ was considered
in \cite {APVV}.
If $b$ and $b^{*}$ are the standard creation and annihilation operators
in the Bosonic Fock space,
$$[b,b^{*}]=1, ~~ b|0>=0,$$
then
\begin{equation}\label{ac}
a=\sqrt{\frac{1-q^{2(N+1)}}{N+1}}b, ~~ c=e^{i\phi}q^{N}
\end{equation}
satisfies the relations (\ref{comcon}). Here $N=b^{*}b, \phi$
is a real number. If $q\to 0$ one gets from (\ref{ac})
\begin{equation}                              \label{acq0}
a=\frac{1}{\sqrt{N+1}}b, ~~ c=e^{i\phi}|0><0|.
\end{equation}

Therefore the master field takes the form
\begin{equation}\label{Phi}
\Phi=b^{*}\frac{1}{\sqrt{N+1}}+\frac{1}{\sqrt{N+1}}b.
\end{equation}

\section{Canonical Master Field Algebra}
\setcounter{equation}{0}

The relations for the master field for D=0 and $g\phi ^{4}$
interaction have the form  \cite {Haan,HalSc}
\begin {equation} 
                                                          \label {cmr}
[\pi, \phi] =-i|0><0|
\end   {equation} 

\begin {equation} 
                                                          \label {vc}
(i\pi +\frac{1}{2}\phi +g \phi ^{3})|0>=0,
\end   {equation} 
where $\phi$  and $\pi$  are Hermitian
\begin {equation} 
                                                          \label {hc}
\phi ^{*}=\phi ,~~\pi ^{*}=\pi
\end   {equation} 
One rewrites them as
\begin {equation} 
                                                          \label {P}
[\pi, \phi] =-iP,
\end   {equation} 
\begin {equation} 
                                                          \label {vcP}
(i\pi +\frac{1}{2}\phi +g \phi ^{3})P=0,
\end   {equation} 

\begin {equation} 
                                                          \label {PP}
P^{2}=P,~~P^{*}=P.
\end   {equation} 
Here we discuss only the case without interaction
\begin {equation} 
                                                          \label {ovc}
(i\pi +\frac{1}{2}\phi)P=0.
\end   {equation} 

Let us define
\begin {equation} 
                                                          \label {ap}
a=\frac{1}{2}\phi +i\pi, ~~ a^{*}=\frac{1}{2}\phi -i\pi.
\end   {equation} 
Then equation (\ref {P}) is equivalent to
\begin {equation} 
                                                          \label {aP}
[a,a^{*}]=P
\end   {equation} 
and (\ref {ovc}) is equivalent to
\begin {equation} 
                                                          \label {VP}
aP=0.
\end   {equation} 

If one has an irreducible representation of (\ref {P}), (\ref {ovc}) or
(\ref {ap}), (\ref {aP})  such
that $P$ is projector on the cyclic vector $|>$
, then there is  a relation
\begin {equation} 
                                                          \label {qv}
\frac{1}{4} \phi ^{2}+\pi ^{2}=1-\frac{1}{2}P,
\end   {equation} 
or equivalently,
\begin {equation} 
                                                          \label {AA}
aa^{*}=1.
\end   {equation} 

Indeed, by acting equation (\ref {aP}) to the vector $|0>$ one gets
$aa^{*}|0>=|0>$. Then $aa^{*}a^{*}|0>=$$(a^{*}a+P)a^{*}|0>=$$a^{*}|0>$,
etc. Therefore for an irreducible representation with a cyclic vector
one has algebras (\ref {P}), (\ref {ovc}), (\ref {qv}) and equivalently
(\ref {aP}), (\ref {VP}), (\ref {AA}). Let us note that from (\ref {aP})
it follows that $P^{*}=P$, and from (\ref {aP}), (\ref {VP}), (\ref {AA})
it follows that $P^{2}=P.$

The coproduct for $\phi ,\pi , P$  elements reads
$$\Delta (\phi ) =\frac{1}{2}\phi \otimes \phi -2 \pi \otimes \pi$$
$$\Delta (\pi ) =\frac{1}{2}(\phi \otimes \pi + \pi \otimes \phi )$$
\begin {equation} 
                                                          \label {app}
\Delta (P ) =P \otimes 1 +  1 \otimes P- P \otimes P
\end   {equation} 
These elements are central.

\section{The Haar Measure on $SU_{q}(2)$ and the Wigner Semicircle
Distribution.}
\setcounter{equation}{0}

The quantum Haar measure $h$ is an
invariant state on the algebra $A$ of functions
on the quantum group, i.e. it satisfies the condition
\begin{equation}
h=(h\otimes id)\Delta=(id\otimes h)\Delta.
\end{equation}
In particular by acting to the
element $g^{i}_{j}$ of the algebra $A_q$ this equality reads
\begin{equation}
h(g^{i}_{j})=\sum_{k=0,1}h(g^{i}_{k}) g^{k}_{j}=
\sum_{k=0,1}g^{i}_{k}h( g^{k}_{j})
\end{equation}
The quantum Haar measure on
compact quantum groups was constructed by Woronovicz  \cite {Wor}
in the following way. One defines the convolution of
two linear functionals $\rho$ and $\chi$ on the Hopf algebra $A$ as
$$(\rho*\chi)(f)=(\rho\otimes\chi)(\Delta (f)), ~~f\in A.$$
One denotes $$\rho^{*k}=\rho *\rho *...*\rho $$
One proves that  for any $f\in A$  there exists the limit
\begin {equation} 
                                                          \label {hm}
h_{\rho}(f)=\lim_{N\to\infty}\frac{1}{N}\sum_{k=1}^{N}\rho^{*k}(f)
\end   {equation} 
that, for the faithful state, does not
depend on the state $\rho$ and defines the quantum Haar measure.
One can interpret (\ref {hm}) as the central limit theorem in
non-commutative probability theory.

By using the bosonization formula (\ref{ac}) for $a$ and $c$ the functional
$h(f)$ can be written as
$$h(f)=\frac{Tr~fe^{-\beta H}}{Tr~e^{-\beta H}};$$
where
$$Tr~f=\sum_{n=0}^{\infty}\frac{1}{2\pi}\int_{0}^{2\pi}<n|f|n>d\phi,$$
$$H=2N,$$
and $|n>$ are $n$-particle oscillator states, $N|n>=n|n>$. Therefore the
quantum Haar functional is the Gibbsian state.
In particular the partition function is
$$Z=Tr~e^{-\beta N}=\frac{1}{1-e^{-2\beta}}.$$
We denote
\begin {equation} 
                                                          \label {int}
\int _{SU_{q}(2)} f d\mu =h(f)
\end   {equation} 

Theory of representations of the quantum group $SU_{q}(2)$
\cite {Wor,VS,Jap}
is similar to the theory
of representations of the classical group $SU(2)$.
An irreducible representation is characterized by its dimension $2l+1, l=0,
\frac{1}{2},1...$. There exists an explicit construction of (2l+1)x(2l+1)
matrix $W^{l}_{km}$, $k,m=-l,...l$, such that
\begin {equation} 
                                                          \label {dw}
\Delta (W_{ij})=\sum _{k}W_{ik}\otimes W_{kj}.
\end   {equation} 
Operators $W^{l}$ satisfy the following orthogonality relations
\begin {equation} 
                                                          \label {WW}
<(W^{j}_{mn})^{*}W^{j'}_{m'n'}>_{q}=\delta _{jj'}\delta _{mm'}\delta _{nn'}
\frac{q^{-2m}}{[2j+1]_{q}}, ~~[n]_{q}=\frac{q^{n}-q^{-n}}{q-q^{-1}}.
\end   {equation} 
The Hopf algebra $A$ of polynomial on $SU_{q}(2))$ has an orthogonal
decomposition (the quantum group analog of the Peter-Weyl theorem)
$$A=\oplus _{l\in N/2}W^{l}$$
with respect to the quantum Haar measure
where $W^{l}$ is spanned by matrix elements
$W^{l}_{mn}$. An element $f$ from $A$ has the Fourier expansion
\begin {equation} 
                                                          \label {qpw}
f=\sum _{l\in N/2}[2j+1]_{q}Tr _{q}(\tilde {f}^{l}W^{l}),
\end   {equation} 
$$\tilde {f}^{l}_{mn}= \int _{SU_{q}(2)} fW_{mn}^{l*} d\mu.$$
For central functions one has a decomposition
over characters
\begin {equation} 
                                                          \label {chd}
f(a+a^{*})=\sum _{l\in N/2}\tilde {f}^{l}\chi _{l}(a+a^{*})
\end   {equation} 
The characters are related with $W$ as
\begin {equation} 
                                                          \label {chdo}
\chi _{l}=\sqrt {\frac{[2l+1]_{q}}{2l+1}}\sum _{n=-l,...l}
q^{n}W^{l}_{nn}
\end   {equation} 
 and there is an explicit formula
\begin {equation} 
                                                          \label {che}
\chi_{l}(t)=\frac{\sin((l+\frac{1}{2})t)}{\sin(\frac{1}{2}t)}=
1+2\cos t +...+2\cos lt;
\end   {equation} 
where $t=2\arccos\frac{a+a^{*}}{2}$.
Note that the explicit form (\ref {che})
 does not depend on the deformation parameter and coincides
with the classical form. There is the recursive relation

\begin {equation} 
                                                          \label {rr}
\chi_{l+1/2}(t)+\chi_{l-1/2}(t)=\chi_{1/2}(t)\chi_{l}(t)
\end   {equation} 
for all $q$ . For $q=0$ this relation gives
\begin {equation} 
                                                          \label {cv}
\chi_{l}(t)|0>=(a^{*})^{2l}|0>.
\end   {equation} 

The orthogonality condition for
characters $\chi_{l}$ also has the same form as in the classical case
\begin {equation} 
                                                          \label {cho}
\int _{SU_{q}(2)}\chi^{*}_{l}(t)\chi_{l'}(t)d\mu=\delta _{ll'}=
\frac{1}{\pi}\int_{0}^{2\pi}\chi^{*}_{l}(\tau)\chi_{l'}(\tau)
(\sin(\frac{\tau}{2}))^{2}d\tau
\end   {equation} 

 From (\ref {chd}) and (\ref {cho})  one notes that to perform the integration
of a polynomial function of the central element $a+a ^{*}$ over quantum
group one can calculate the integral over classical group of the same
polynomial,

\begin {equation} 
                                                          \label {pi}
\int _{SU_{q}(2)}f(t)d\mu=
\frac{1}{\pi}\int_{0}^{2\pi}f(\tau)(\sin(\frac{\tau}{2}))^{2}d\tau ,
\end   {equation} 
or
\begin {equation} 
                                                          \label {pi'}
\int _{SU_{q}(2)}f(a+a^{*})d\mu=
\frac{1}{2\pi}\int_{-2}^{2}f(\lambda)\sqrt{4-\lambda^{2}}d\lambda .
\end   {equation} 
One gets the Wigner semicircle distribution for any $q$.
In particular for $q\to 0$ in the left hand side of (\ref {pi'}) one has
\begin {equation} 
                                                          \label {wd}
\int _{SU_{0}(2)}(a+a^{*})^{k}d\mu=<0|(a+a^{*})^{k}|0>
\end   {equation} 
and therefore
 \begin {equation} 
                                                          \label {wd'}
<0|(a+a^{*})^{k}|0>=
\frac{1}{2\pi}\int_{-2}^{2}\lambda^{k}
\sqrt{4-\lambda^{2}}d\lambda.
\end   {equation} 
that demonstrates the well-known
Wigner distribution  for the master field  from quantum group point of view.

Note that for the case $q=0$ the set of central functions is more large as
compare with the case of arbitrary $q$. It is spanned by functions on $a$
and $a^{*}$. For these central functions one can write
down the special form of the Peter-Weyl decomposition.

The relation between  the master field algebra and the quantum group
$SU_{q}(2)$ permits to write immediately the
coherent states for master field as well as
for operators  $a$ and $a^{*}$. Coherent states for
$SU_{q}(2)$   have the form \cite {APVV}
\begin {equation} 
                                                          \label {cs}
\Psi (u)=\sum _{j\in N/2}\sqrt {(2j+1)[2j+1]_{q}}tr _{q}(W^{j*}T^{j}(u))
\end   {equation} 
Here $u$ is an element of $SU(2)$,
$$g=\left(\begin{array}{cc}\alpha &-\beta^{*}\\\beta&\alpha ^{*}
\end{array}\right)$$
and $$T^{j}(u)=(D^{j}_{mn}(u))$$
is a unitary representation of $SU(2)$ of spin $j$. From (\ref {cs}) and
(\ref {WW}) it follows that if we introduce the kernel
 by
\begin {equation} 
                                                          \label {fa}
{\cal K}(u,u')=\int _{SU_{q}(2)}\bar{\Psi}(u)\Psi (u') d\mu
\end   {equation} 
then this kernel satisfies the superposition relation
\begin {equation} 
                                                          \label {sr}
\int {\cal K}(u,u'){\cal K}(u',u'')du'={\cal K}(u,u''),
\end   {equation} 
where $du$ stands for the Haar measure on $SU(2)$.

One can also introduce the coherent states for the master field $a+a^{*}$
\begin {equation} 
                                                          \label {sscs}
\Psi _{c} (\alpha +\alpha ^{*})=\sum _{j\in N/2}(\chi _{j} (a+a^{*}))^{*}
\chi _{j}
(\alpha +\alpha ^{*})
\end   {equation} 

  Introducing the kernel
\begin {equation} 
                                                          \label {fac}
{\cal K}(\alpha+\alpha^{*},\alpha '+\alpha ^{*'})=
\int _{SU_{q}(2)}\bar{\Psi_{c}}(\alpha
+\alpha ^{*})
\Psi _{c}(\alpha '+\alpha '^{*}) d\mu
\end   {equation} 
 we get the superposition property
\begin {equation} 
                                                          \label {ssr}
\frac{1}{\pi}\int_{0}^{2\pi}
 {\cal K}(\tau,\tau'){\cal K}(\tau',\tau'')(\sin(\frac{\tau '}{2}))^{2}d\tau'=
 {\cal K}(\tau,\tau''),
\end   {equation} 
$\tau= 2\arccos\frac{\alpha +\alpha ^{*}}{2}$

 From (\ref {cv}) for $q=0$ we have
\begin {equation} 
                                                          \label {csv}
\Psi _{c}(\alpha +\alpha ^{*})|0>=
\sum _{j\in N/2} \chi _{j}
(\alpha +\alpha ^{*}) (a^{*})^{2j}|0>
\end   {equation} 
and
\begin {equation} 
                                                          \label {K0}
{\cal K}(\alpha+\alpha^{*},\alpha '+\alpha ^{*'})=
\sum _{j\in N/2}(\chi _{j} (\alpha+\alpha^{*}))^{*}
\chi _{j} (\alpha '+\alpha '^{*})
\end   {equation} 

In conclusion, in this paper we discussed a relation between
the simplest master field and quantum group $SU_q(2)$. It would
be interesting to extend such a relation to more general
master fields and quantum groups.
$$~$$
$$~$$
{\bf ACKNOWLEDGMENT}
$$~$$
I.A. and I.V. are supported in part  by
International Science Foundation under the grant M1L000.
I.V. thanks Centro  V. Volterra   Universita di Roma Tor Vergata
where this work was started for the
 kind hospitality.
I.A. thanks G.Arutyunov, P.Medvedev and A.Zubarev for useful discussions.

$$~$$

\end{document}